# The VST Early-type GAlaxy Survey: Exploring the Outskirts and Intra-cluster Regions of Galaxies in the Low-surface-brightness Regime

Enrichetta Iodice[1]
Marilena Spavone[1]
Massimo Capaccioli[1]
Pietro Schipani[1]
Magda Arnaboldi[2]
Michele Cantiello[3]
Giuseppe D'Ago[4]
Demetra De Cicco[4]
Duncan A. Forbes[5]
Laura Greggio[6]
Davor Krajnović[7]
Antonio La Marca[1,8]
Nicola R. Napolitano[9]
Maurizio Paolillo[8]
Rossella Ragusa[1,8]
Maria Angela Raj[10]
Roberto Rampazzo[11]
Marina Rejkuba[2]

[1] INAF–Astronomical Observatory of Capodimonte, Naples, Italy
[2] ESO
[3] INAF–Astronomical Abruzzo Observatory, Teramo, Italy
[4] Institute of Astrophysics, Pontificia Universidad Católica de Chile, Santiago, Chile
[5] Centre for Astrophysics and Supercomputing, Swinburne University of Technology, Hawthorn, Australia
[6] INAF–Astronomical Observatory of Padova, Padova, Italy
[7] Leibniz Institute for Astrophysics Potsdam, Germany
[8] University of Naples, Italy
[9] School of Physics and Astronomy, Sun Yat-sen University Zhuhai Campus, Guangdong, China
[10] INAF–Astronomical Observatory of Rome, Italy
[11] INAF–Astronomical Observatory of Padova, Asiago, Italy

**The VST Early-type GAlaxy Survey[1] (VEGAS) is a deep, multi-band (*u*, *g*, *r*, *i*) imaging survey, carried out with the 2.6-metre VLT Survey Telescope (VST) at ESO's Paranal Observatory in Chile. VEGAS combines the wide (1-square-degree) field of view of the VST's OmegaCAM imager and long integration times, together with a specially designed observing strategy. It has proven to be a gold mine for studies of features at very low surface brightness, down to levels of $\mu_g$ ~ 27–30 magnitudes arcsec$^{-2}$, over 5–8 magnitudes fainter than the dark sky at Paranal. In this article we highlight the main science results obtained with VEGAS observations of galaxies across different environments, from dense clusters of galaxies to unexplored poor groups and in the field.**

## VEGAS in the panorama of deep imaging surveys

Exploring the low-surface-brightness (LSB) Universe is one of the most challenging tasks in contemporary astrophysics. It is important for mapping the mass assembly of galaxies in all environments and thus constraining their formation within the Lambda Cold Dark Matter (ΛCDM) paradigm. In this framework, clusters of galaxies are expected to grow over time by accreting smaller groups along filaments, driven by the effect of gravity generated by the total matter content. In the deep potential well at the cluster centre, the galaxies continue to undergo active mass assembly. As a result, gravitational interactions and merging between systems of comparable mass and/or smaller objects play a fundamental role in defining the galaxies' morphology, the build-up of the stellar halos, and the intra-cluster light (ICL). Stellar halos are extended (≥ 100 kpc) and faint ($\mu_g$ ≥ 26–27 magnitudes arcsec$^{-2}$) components made of stars stripped from satellite galaxies, in the form of streams and tidal tails, with multiple stellar components and complex kinematics (see the review by Duc, 2017). The ICL forms during the infall of groups of galaxies into the cluster potential as material is stripped from the galaxies' outskirts. This diffuse and very faint component ($\mu_g$ ≥ 28 magnitudes arcsec$^{-2}$) grows over time with the mass assembly of the cluster, to which the relics of the interactions between galaxies (stellar streams and tidal tails) also contribute.

In the last two decades, deep imaging surveys have given a huge boost to the study of mass assembly in different environments by providing extensive analyses of the light and colour distributions of galaxies out to the regions of stellar halos and the intra-group/intra-cluster space (see the review by Mihos, 2019). Investigations of mass assembly in the outskirts of galaxies have also been conducted by means of stellar kinematics and population properties of discrete tracers like globular clusters (GCs) and planetary nebulae.

The main goal of imaging and spectroscopy surveys is to provide a set of observables that can be directly compared with detailed theoretical models of the structure and stellar populations of stellar halos, the formation of the ICL and the amount of substructure in various environments (see the review by Arnaboldi et al., 2020). In this context, VEGAS has played a pivotal role in exploring the properties of galaxies as a function of environment down to the LSB regime. To date, using about 400 hours of observing time, VEGAS has already collected data on 35 groups and clusters of galaxies, covering a total area on the sky of ~ 70 square degrees. About 30% of the VEGAS observing time was dedicated to the Fornax Deep Survey (FDS; Peletier et al., 2020). The FDS covers the Fornax cluster out to the virial radius (~ 0.7 Mpc), taking in an area of 26 square degrees around the central galaxy NGC 1399 and including the SW subgroup centred on NGC 1316.

Based on the analysed data, VEGAS and the FDS have allowed us to a) study the outskirts of the galaxies and detect the ICL and LSB features in the intra-cluster/group space (see Iodice et al., 2016; Spavone et al., 2018; Iodice et al., 2020a and references therein), b) trace the mass assembly process in galaxies by estimating the accreted mass fraction in the stellar halos and provide results that can be directly compared with the predictions of galaxy formation models (see Spavone et al. 2020), c) trace the spatial distribution of candidate GCs (see Cantiello et al., 2020 and references therein), d) provide the largest size- and magnitude-limited catalogue of dwarf galaxies in the Fornax cluster (Venhola et al., 2018), and e) detect ultra-diffuse galaxies (UDGs; Forbes et al., 2020; Iodice et al., 2020b).

With the first data release (DR1) of VEGAS we have provided the reduced VST mosaics of 10 targets, recently published by the VEGAS collaborations (Capaccioli et al., 2015; Spavone et al., 2017, 2018; Iodice et al., 2020a; Cantiello et al., 2018). The data products (i.e., images in all observed bands) are available via the ESO Science Portal[2].

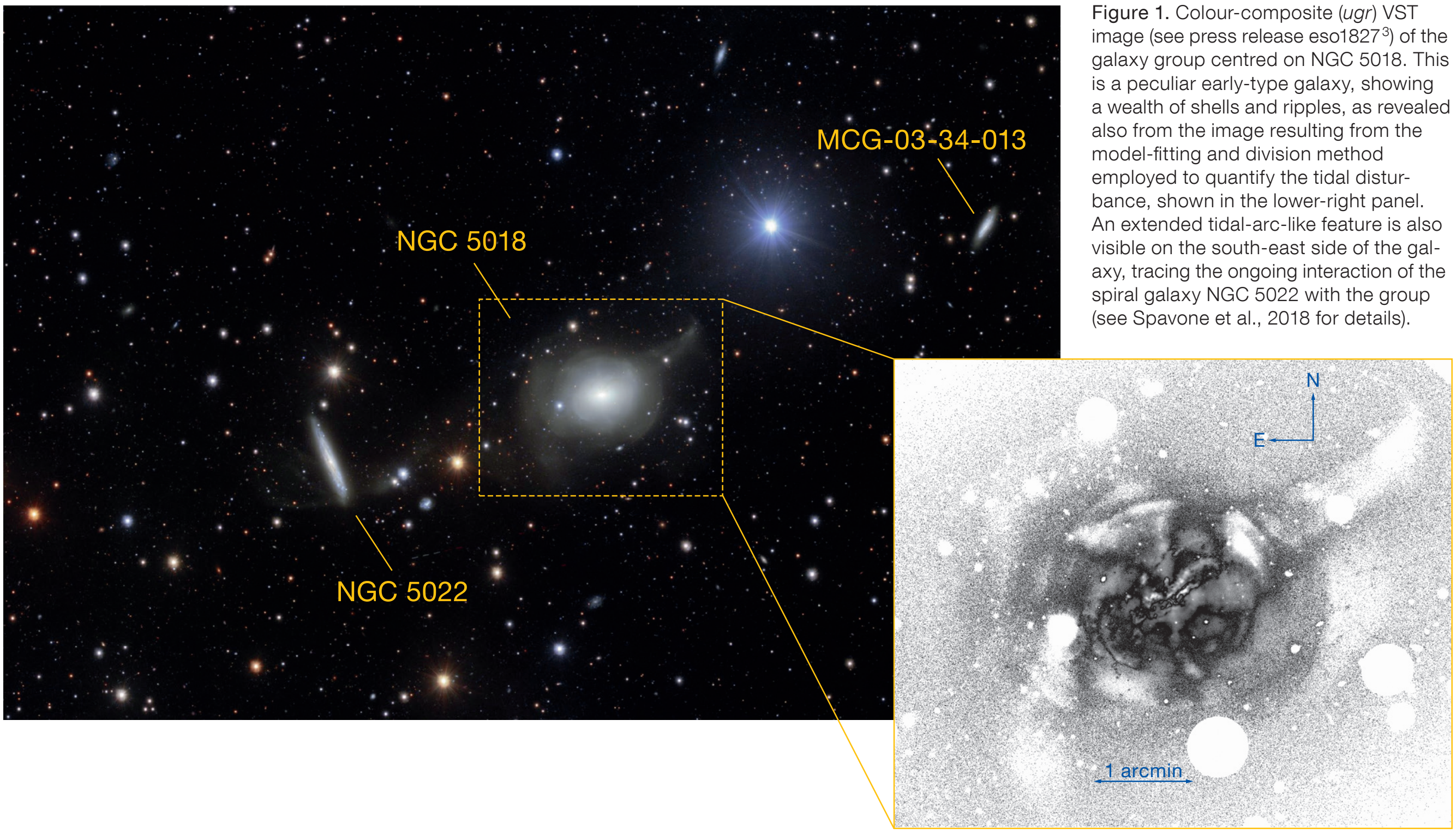

Figure 1. Colour-composite (*ugr*) VST image (see press release eso1827[3]) of the galaxy group centred on NGC 5018. This is a peculiar early-type galaxy, showing a wealth of shells and ripples, as revealed also from the image resulting from the model-fitting and division method employed to quantify the tidal disturbance, shown in the lower-right panel. An extended tidal-arc-like feature is also visible on the south-east side of the galaxy, tracing the ongoing interaction of the spiral galaxy NGC 5022 with the group (see Spavone et al., 2018 for details).

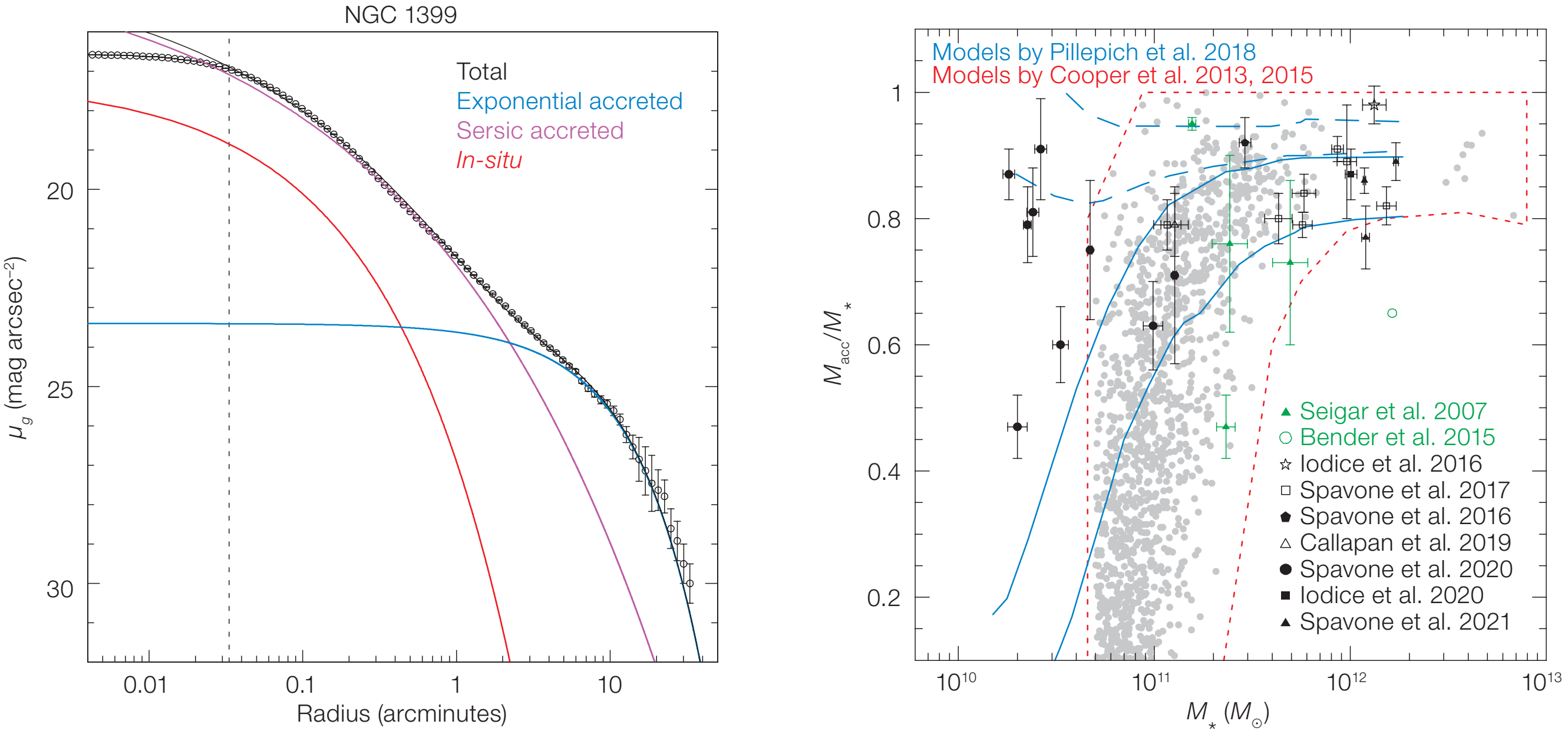

Figure 2. Left panel: VST *g*-band profile of NGC 1399 (from the FDS, Iodice et al., 2016: see also press release eso1827[4]) fitted with a three-component model motivated by the predictions of theoretical simulations. The total accreted mass fraction (shown in the right panel) is derived from the contribution of the Sérsic accreted and exponential accreted components with respect to the total luminosity of the galaxy (see Spavone et al., 2020 for details). Right panel: Accreted mass fraction ($M_{acc}/M_*$) versus total stellar mass ($M_*$) for early-type galaxies. Green points correspond to the brightest cluster galaxies from the literature. Black symbols are for VEGAS galaxies. The region outlined by the red dotted line encloses the predictions of cosmological galaxy formation simulations, which are indicated as grey dots. Blue continuous and dashed lines indicate the accreted mass fraction measured within 30 kpc and outside 100 kpc, respectively, in Illustris simulations. For further details, see Spavone et al. (2020) and reference therein.

## Methods: what observables can be derived from deep imaging data?

To trace mass assembly on all scales, i.e., in galaxies and clusters, we need to derive observables that can be directly compared with theoretical predictions. They are listed below.

– Galaxy morphology: in order to detect any asymmetry in the outskirts of a galaxy and the remnants of past accretion/merging events (such as tidal tails, stellar streams and shells) and also the ICL, a 2D model of the light distribution is derived by fitting the isophotes and it is then subtracted from the parent image. An example is given in Figure 1. All steps are described in detail in VEGAS papers (for example, Capaccioli et al., 2015; Iodice et al., 2016).
– Azimuthally averaged surface brightness (SB) profiles: these are derived from the isophote fit and, by performing a 1D fit using multi-component empirical laws, they are used to set the scales of the different components in the galaxy that dominate the light distribution, i.e., the stellar envelope and ICL versus central *in-situ* stars. The fitting algorithm and tools are presented in Spavone et al. (2020) and references therein. An example is shown in Figure 2.
– Colour gradients: average colour profiles are derived from the SB profiles in order to address the contribution of different stellar populations in the outskirts of the galaxies (Spavone et al., 2020).
– Inventories of GCs, dwarf galaxies and UDGs: in the intra-cluster regions, by using automatic detection tools, we can map the number density and structural properties of the small stellar systems, since they are the main contributors to, and are prominent tracers of, the build-up of the stellar halos and the ICL. An extensive description of the strategy and tools used for the detection and analysis, based on VST data, of the GCs and dwarf galaxies is provided by Cantiello et al. (2020) and Venhola et al. (2018), respectively. The 2D density map of GCs in the Fornax cluster obtained by Cantiello et al. (2020) is shown in Figure 3.

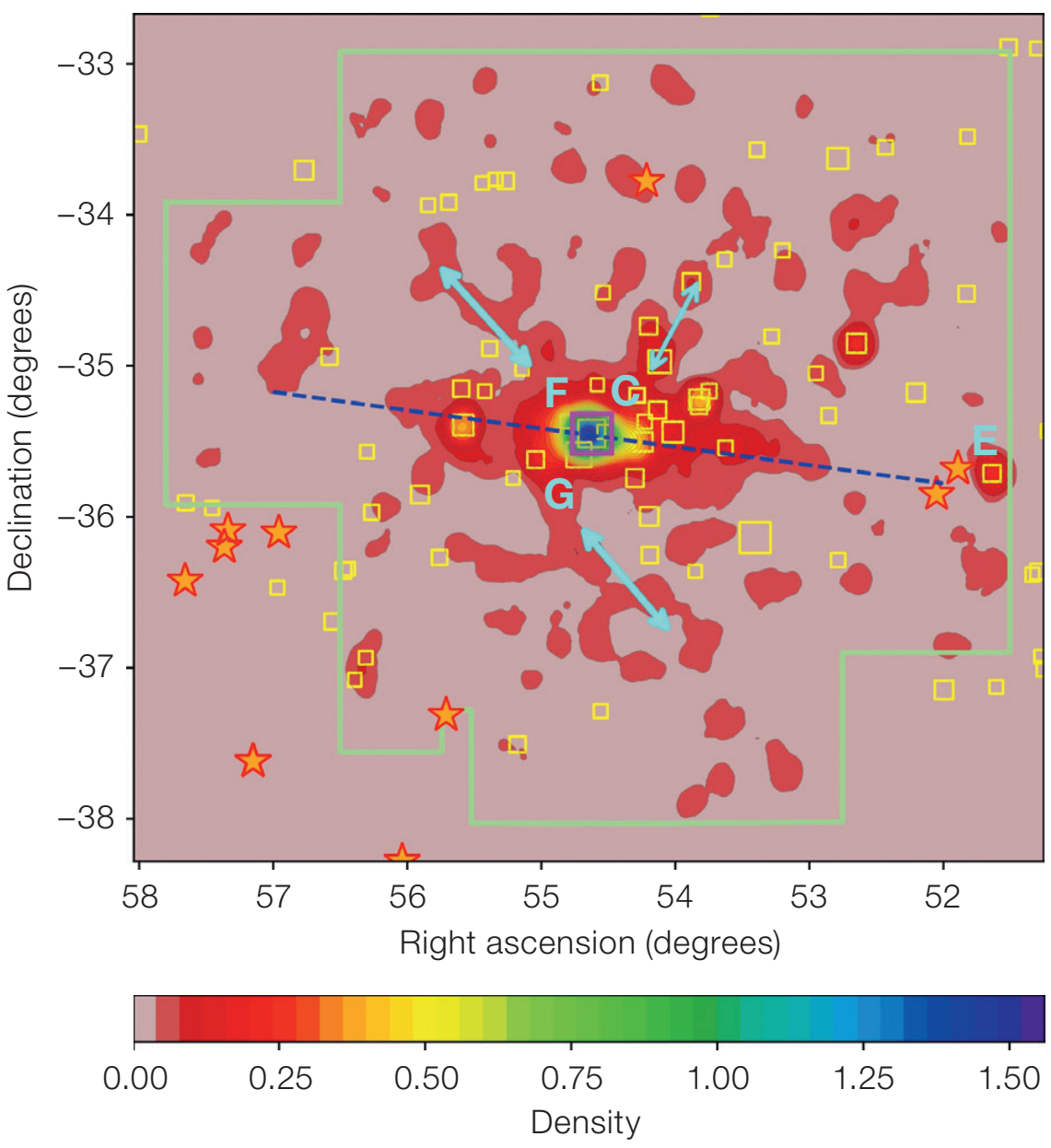

Figure 3. Single-panel view of the 2D globular clusters surface distribution over the FDS area by Cantiello et al. (2020). The density is in number of candidates per square arcminute. East is left, north is up. The light green line shows the FDS footprint. Five-pointed stars mark stars with $m_V \leq 7$ mag; yellow squares show galaxies brighter than $B_T = 16$ mag, with symbol size scaled to galaxy total magnitude; NGC 1399 is marked with a magenta square. Light blue arrows and the labels C, F, G indicate the globular cluster overdensities found in the Fornax cluster area. The blue dashed line shows the ~ 10-degree tilt in the direction of NGC 1336 (labelled E in the figure).

## Outcomes: observations versus theoretical predictions

Combining the observables derived from the VST data with theoretical models of the structure and stellar populations of stellar halos, ICL formation and the amount of substructure in various kinds of environment leads to the main results of VEGAS, as summarised below.
– Based on the detection and characterisation of the fine structures and average colours in the outskirts of galaxies, we have been able to trace the **formation history of the stellar halo** (see Iodice et al., 2017; Spavone et al., 2018; Iodice et al., 2020a and references therein). Figure 4 shows some examples of the brightest galaxy in the centre of a group, showing that the outer envelope still hosts the remnants of the accreted satellite galaxies that are forming the stellar halo. For these objects, we examined possible formation scenarios by comparing the observed properties (morphology, colours, gas content) with predictions from cosmological simulations of galaxy formation. This allowed us to address the formation of stellar halos from the accretion of small satellites or as the result of past major merging events.
– Based on 1D fitting of the SB profiles, we have provided an **estimate of the accreted mass fraction** in the brightest and most massive galaxies. The right panel of Figure 2 shows the accreted mass fraction estimated for the most massive galaxies in groups and clusters from VST deep images, recently published by Spavone et al. (2020). This represents one of the major achievements of VEGAS, as it is the first attempt to compare this quantity with theoretical predictions of mass assembly as a function of the total stellar mass in galaxies. The results suggest that, in agreement with theoretical models, the largest accreted mass fraction is found in the most massive galaxies.
– We have derived the **ICL fraction in several groups and clusters**. This is related to the look back time of the mass assembly, since for more evolved structures (groups or clusters of galaxies) we expect a larger amount of ICL. In Figure 5 we show all the ICL estimates we have derived for the analysed VEGAS targets. In accordance with simulations, the ICL fraction ranges from 10% to 45% in massive ($\geq 10^{12}\ M_{\odot}$) groups or clusters of galaxies. The lower-mass regime remains quite unexplored, from both the observational and theoretical sides.
– By studying the structure and colours of late-type disc galaxies, we have addressed the role of the environment in driving the **evolution of galaxies** (see Raj et al., 2020 and references therein).

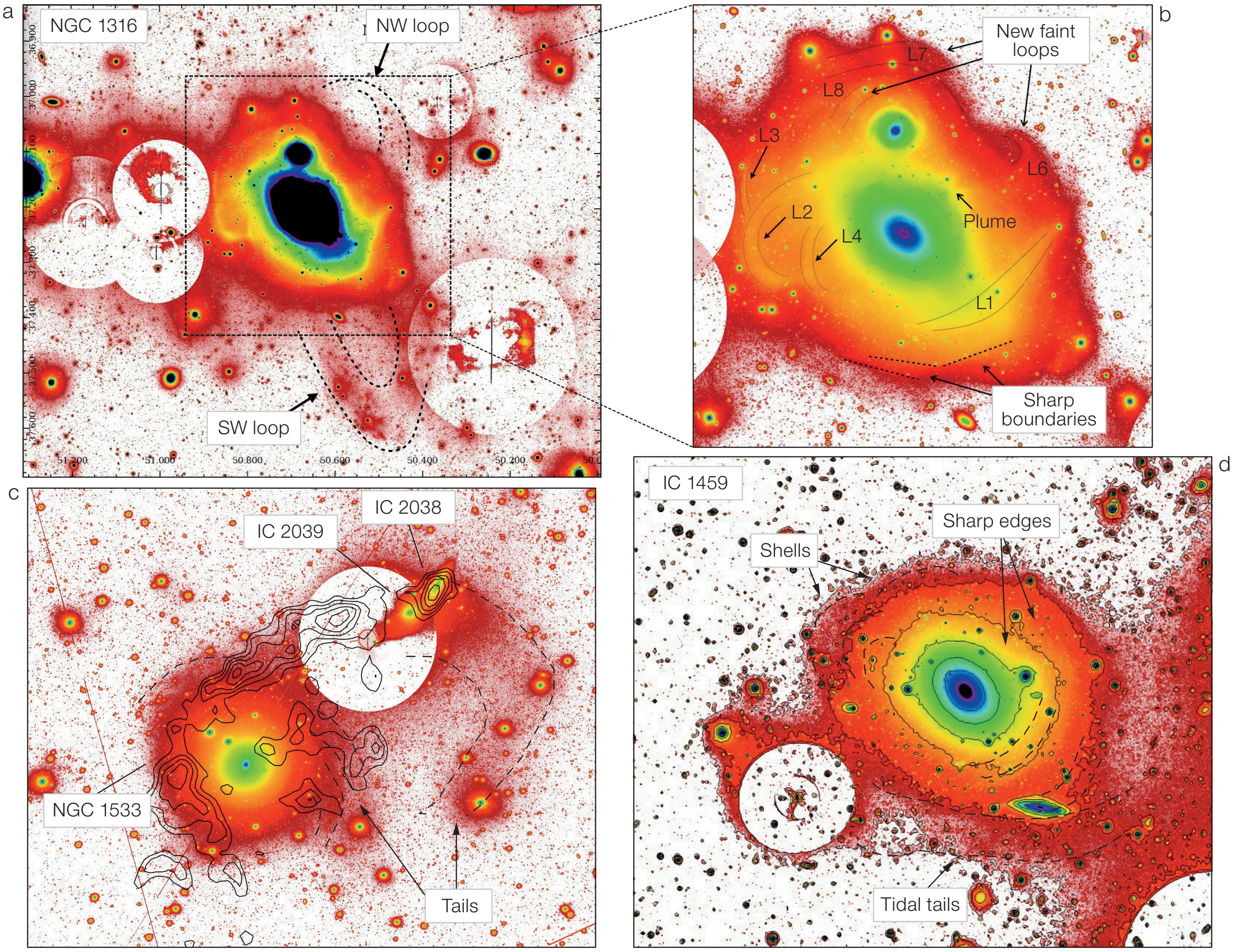

Figure 4. (Above) Galaxies from VEGAS showing the remnants of accreted satellite galaxies in their outskirts (see Iodice et al., 2020a and references therein). Panel a shows the *g*-band image of NGC 1316[5], the brightest galaxy in the centre of the SW subgroup in the Fornax cluster. The giant (160 kpc) SW loop and the faint NW loop (90 kpc) are marked with long dashed lines. The enlarged region in panel b shows the wealth of loops and shells at smaller radii in NGC 1316, which are probably due to the disruption of dwarf galaxies (see Iodice et al., 2017). Panel c shows the *g*-band image of NGC 1533, the brightest of the triplet of galaxies in the Dorado group. The H I map from the Australia Telescope Compact Array (ATCA) is superimposed (black contours). The faint stellar tails, which trace the accretion in the outskirts, are marked as dashed black lines. Panel d shows the *g*-band image of IC 1459, the brightest group member, where the shells and tails in the outskirts are signs of ongoing accretion and interactions.

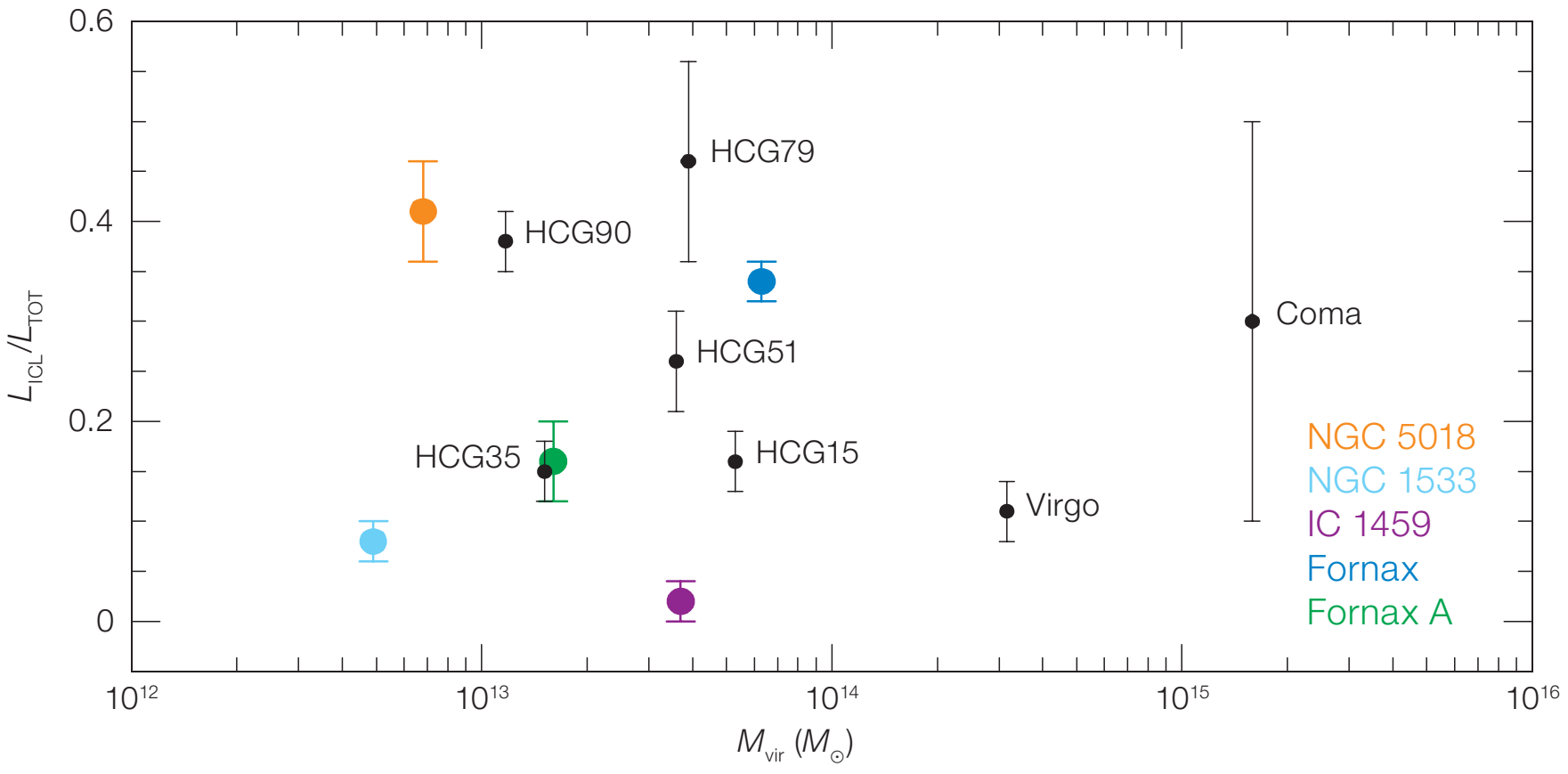

Figure 5. Intra-cluster light (ICL) fraction as a function of halo mass for VEGAS targets and for the Fornax cluster from the FDS. These are compared with other measurements for the Virgo and Coma clusters. Values are also compared with those for several Hickson Compact groups (HCGs) (see references in Iodice et al., 2020a and Spavone et al., 2020).

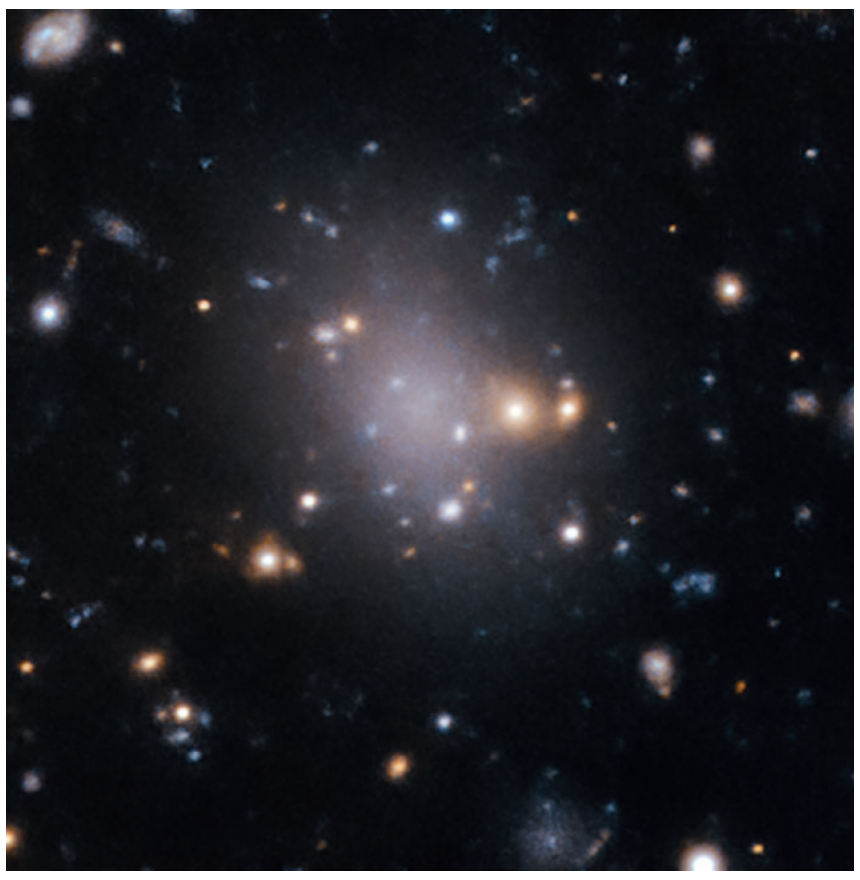

Figure 6. Colour-composite image of a UDG candidate detected in the Hydra I cluster from the deep VST images (see Iodice et al., 2020b). The image size is about 16 × 16 kpc.

## Ultra-diffuse galaxies

UDGs have a special role in the realm of the LSB Universe, since they are amongst the faintest bound systems in groups and clusters of galaxies. UDGs are empirically defined to be faint ($\mu_{0,g} \geq 24$ magnitudes arcsec$^{-2}$) and diffuse ($R_e \geq 1.5$ kpc) objects, with stellar masses similar to those of dwarf galaxies ($10^7$–$10^8$ $M_\odot$; van Dokkum et al., 2015). They constitute the extreme tail of the size-luminosity distribution of dwarf galaxies.

Although the detection and analysis of UDGs are challenging, owing to their LSB nature, a significant population of UDGs has been found in dense environments such as clusters and groups of galaxies as well as in the field (for example, Janssens et al., 2019). The discovery of a number of UDGs with very low dark matter content raised new questions about the whole framework of galaxy formation in order to account for such long-lived, large and baryon-dominated stellar systems (van Dokkum et al., 2016).

Using FDS and VEGAS data, UDGs were discovered in groups and clusters of galaxies (see Prole et al., 2019; Forbes et al., 2020 and references therein). In particular, Iodice et al. (2020b) presented the first sample of UDG candidates in the Hydra I cluster (see Figure 6). For each UDG, we analysed the light and colour distribution, estimated the stellar mass, and provided a census of the GC systems around it. Based on the GC populations of these newly discovered UDGs, we conclude that most of these galaxies have a standard or low dark matter content, with a halo mass of $\leq 10^{10}$ $M_\odot$, comparable to dwarf galaxies of similar stellar masses. These results represent an important step in our project to enlarge the number of confirmed UDGs.

## Future perspectives

By the end of the survey (2022), VEGAS will have collected a total of 55 targets, with a spatial coverage of ~ 90 square degrees and spanning a halo mass range of $10^{10}$–$10^{14}$ $M_\odot$. With such a large dataset we have started two new projects: the study of the lowest stellar mass regime ($\leq 10^{10}$–$10^{12}$ $M_\odot$), typical of the low-density environments, such as groups of galaxies, which allows us to fill the gap in the comparison with the theoretical predictions across this range of stellar mass; and the detection and analysis of a large number (~ 1300) of UDGs using the entire VEGAS sample, named Ultra-VEGAS, which will provide a statistically relevant UDG sample with which to constrain the nature and formation of these systems.

In addition, for the extension of VST operations beyond 2022 and in the footsteps of VEGAS, we have proposed a deep multi-band imaging survey that aims to map the large-scale structure of two superclusters in the nearby Universe: the Fornax-Eridanus and the Hydra I-Centaurus superclusters. This project, named VEGAS-LSS, aims to exploit the excellent photometric wide-field capabilities of the VST to study the unexplored regions of voids and filaments in the large-scale structure around groups and clusters down to the LSB regime. Based on the uniform deep survey of the large-scale structure, the main science goal of VEGAS-LSS is to provide the key observables needed to trace the infall of groups into clusters and, thereby, important clues as to the clusters' assembly histories. The proposed projects will give a timely and valuable return on the large effort dedicated by the astronomers and engineers of INAF, University of Naples and ESO, to building and maintaining the VST facility, as well as to the international astronomical community in planning and pursuing innovative science at the threshold of future survey telescopes.

## Acknowledgements

We would like to acknowledge all the scientists who contributed to the VEGAS and FDS projects: E. Iodice (PI), M. Spavone (co-PI), M. Capaccioli (former PI), M. Arnaboldi, E. Bannikova, D. Bettoni, M. den Brok, S. Brough, M. Cantiello, A. Cattapan, S. Ciroi, A. P. Cooper, E. M. Corsini, R. D'Abrusco, G. D'Ago, E. Dalla Bontà, D. De Cicco, E. Emsellem, J. Falcon-Barroso (FDS co-I), D. Forbes, A. Grado, L. Greggio, M. Gullieuszik, E. Held, M. Hilker, D. Krajnović, A. La Marca, L. Limatola, S. Mieske, N. R. Napolitano, M. Paolillo, A. Pasquali, R. Peletier (FDS co-PI), A. Pizzella, I. Prandoni, R. Ragusa, M. A. Raj, M. Rejkuba, R. Rampazzo, P. Schipani (VST PI), C. Spiniello, G. van de Ven (FDS co-I).

The authors acknowledge financial support from the VST project (PI: P. Schipani). E. Iodice acknowledges financial support from ESO during a science visit to the Garching Headquarters from September 2019 to August 2020.

## Links

[1] VEGAS survey website: https://www.na.astro.it/vegas/VEGAS/Welcome.html
[2] VEGAS first data release acess via the ESO Archive Science Portal: https://www.eso.org/sci/observing/phase3/news.html#VEGAS-DR1
[3] ESO press release eso1827: https://www.eso.org/public/news/eso1827/
[4] ESO press release eso1612: https://www.eso.org/public/news/eso1612/
[5] ESO press release eso1734: https://www.eso.org/public/news/eso1734/